\documentclass[twocolumn]{article}
\usepackage{amsmath,epsfig}

\usepackage{amsmath}
\usepackage{bm}
\usepackage{url}
\usepackage{svg}
\usepackage{tikz, graphicx}
\usepackage{standalone}
\usepackage{hyperref}
\usepackage{arydshln}
\usepackage{csquotes}
\usepackage{cite}

\newcommand\scalemath[2]{\scalebox{#1}{\mbox{\ensuremath{\displaystyle #2}}}}

\usepackage{enumitem}
\usepackage[T1]{fontenc}

\title{Robust Rational Polynomial Camera Modelling \\
for SAR and Pushbroom Imaging}

\author{Roland Akiki$^{1,2}$\hspace{0.45 cm} Roger Mar\'{i} $^{1}$  \hspace{0.45 cm} Carlo de Franchis$^{1,2}$\\Jean-Michel Morel$^1$ \hspace{0.45 cm} Gabriele Facciolo$^1$}
\date{$^1$ Université Paris-Saclay, CNRS, ENS Paris-Saclay, Centre Borelli, France\\$^2$ Kayrros SAS}

\begin{document}
\maketitle
\begin{abstract}
The Rational Polynomial Camera (RPC) model can be used to describe a variety of image acquisition systems in remote sensing, notably optical and Synthetic Aperture Radar (SAR) sensors. RPC functions relate 3D to 2D coordinates and vice versa, regardless of physical sensor specificities, which has made them an essential tool to harness satellite images in a generic way. This article describes a terrain-independent algorithm to accurately derive a RPC model from a set of 3D-2D point correspondences based on a regularized least squares fit. The performance of the method is assessed by varying the point correspondences and the size of the area that they cover. We test the algorithm on SAR and optical data, to derive RPCs from physical sensor models or from other RPC models after composition with corrective functions.
\end{abstract}
\section{Introduction}
\label{sec:intro}

Developing a remote sensing application requires a set of tools, one of which is geolocation. Geolocation relates the 3D world coordinates to the 2D image. This is represented by means of a \textit{projection} function $\mathcal{P}: \mathbf{R}^3 \rightarrow \mathbf{R}^2 $, that maps 3D points to the image plane, and its inverse, the \textit{localization} function $\mathcal{L}: \mathbf{R}^2 \times \mathbf{R} \rightarrow \mathbf{R}^3 $. When all the physical phenomena and components involved in the acquisition process are known, the geolocation functions can be defined by a chain of operations that model such factors, in what is known as a \textit{physical} or \textit{rigorous} sensor model.

Pushbroom scanners are the most common optical satellite image acquisition system, typically consisting of a single line of pixel sensors mounted on a platform that captures each line of the image at a different moment in time. As a result, the exterior orientation parameters, i.e. the perspective center and the attitude angles, change from line to line. The intrinsic parameters (e.g. pixel size, focal length, lens distortion), related to the physical design of the sensor, are constant across the image \cite{grodecki2001ikonos}. A detailed description of a simplified physical sensor model for pushbroom scanners can be found in \cite{de2015attitude}. 
\begin{figure}[t]
\centering
 \includegraphics[width=0.53\textwidth]{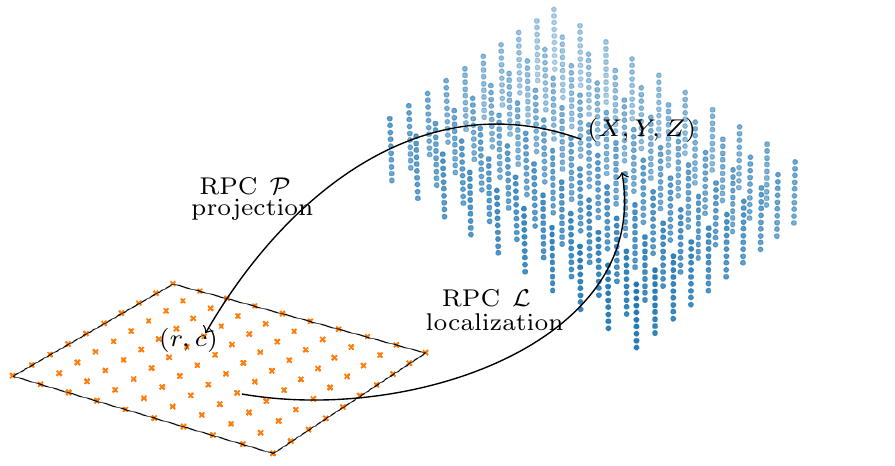}
 \caption[]{The RPC model is derived using a grid of 3D CNPs (Control Points) and its projection onto the satellite image.}
\label{fig:teaser}
\end{figure}
In the case of Synthetic Aperture Radar (SAR) images, the most used physical sensor model is the Range-Doppler model detailed in \cite{curlander1982location}. SAR satellites send an electromagnetic wave that is reflected on the ground. The image is acquired line by line (similar to a pushbroom system), and the position of a ground patch in the image is related to its distance to the sensor, known as the \textit{range}. The Range-Doppler model is constructed based on ephemeris data (time, position and velocity samples along the orbit) and the acquisition timing information. The ephemeris data needs to be interpolated to obtain continuous geolocation functions along the orbit. 

Image vendors have adopted the generic Rational Polynomial Camera\footnote{Also referred to as Rational Polynomial Coefficients camera model or Rational Function Model in the literature.} (RPC) model to save customers from having to deal with the complex specificities of rigorous sensor models. The RPC model is independent of physical properties and offers flexibility to work with different coordinate systems. RPCs have become essential metadata to process satellite images in a generic way, from different sources and for multiple tasks, e.g. photogrammetry and radargrammetry based 3D reconstruction or image ortho-rectification and coregistration. 

In this article, we describe a terrain-independent algorithm to fit a RPC model from a physical sensor model or any other geolocation model. Our contributions are:
\begin{itemize}
    \item[-] An open-source implementation of the method as an easy-to-use Python package, which is available at  {\small{\url{https://github.com/cmla/rpcfit}}}.

    \item[-] An evaluation of the algorithm's precision and robustness based on real scenarios. We test our method using Sentinel-1 and WorldView-3 images, to fit a SAR physical sensor model or correct an existing RPC model by composing it with a complementary transformation.
\end{itemize}

\subsection{Mathematical formulation of the RPC model}
\label{sec:rpc_maths}

The RPC model defines the projection function $\mathcal{P}$ as
\begin{equation}
r_{n}=\frac{a\left(X_{n}, Y_{n}, Z_{n}\right)}{b\left(X_{n}, Y_{n}, Z_{n}\right)} \quad c_{n}=\frac{e\left(X_{n}, Y_{n}, Z_{n}\right)}{f\left(X_{n}, Y_{n}, Z_{n}\right)},
\label{eq:rpc_projection}
\end{equation}
where $a, b, e, f$ are cubic polynomials. $X, Y, Z$ represent the longitude, latitude and height of a 3D point; and $r, c$ are the row and column of its projection on the image plane.

Equation~\ref{eq:rpc_projection} uses normalized coordinates for better numerical stability, hence the subscript $n$. Normalized values are in the range [-1, 1] and they are obtained using two scalars, an offset and a scale factor: $X_{n}=(X-X_{\text{offset}})/X_{\text {scale}}$, 
where $X$ could be $r, c, X, Y$ or $Z$ from Equation~\ref{eq:rpc_projection}.

Each RPC polynomial $p$ is defined by 20 coefficients as
\begin{equation}
\scalemath{0.9}{
\textstyle    \begin{aligned}
p\left(X, Y, Z\right) &= \, \, p_0 + p_1Z + p_2Y + p_3X + p_4ZY + p_5ZX \\ 
 +& p_6YX + p_7X^2 + p_8Y^2 + p_9Z^2 + p_{10}ZYX \\
+& p_{11}Z^2Y + p_{12}Z^2X + p_{13}Y^2Z + p_{14}Y^2X \\
+& p_{15}ZX^2 + p_{16}YX^2 + p_{17}Z^3 + p_{18}Y^3 + p_{19}X^3,
    \end{aligned}}
\end{equation}
where $p_{i}$ is the $i$-th coefficient of $p$. Since we set $p_0=1$ for the RPC denominator polynomials, a total of 78 coefficients need to be determined to define $a$, $b$, $e$ and $f$ in Equation~\ref{eq:rpc_projection}.
\section{Related work}
\label{sec:related_work}

RPCs have been used for high-resolution optical satellite imaging since the launch of Ikonos in 1999 \cite{grodecki2001ikonos, tao2001comprehensive, fraser2006sensor, long2015rpc}. In the last decade, they have been proven to be extremely accurate for SAR acquisition systems as well \cite{zhang2010evaluation, zhang2011rational}. 

The RPC model of a satellite image can be constructed using a set of correspondences between image and object space coordinates. Depending on the nature of these correspondences, the literature can be classified into terrain-dependent or \mbox{independent} methods (or a combination of both). Terrain-dependent strategies use Ground Control Points (GCPs), whose object and image coordinates are known in advance relying on manual labeling or on-site measurements. Oppositely, terrain-independent methods derive virtual sets of 2D-3D point correspondences from other \mbox{geolocation} functions, usually a physical sensor model. Once the point correspondences are available, least squares algorithms are typically used to estimate the RPC coefficients that minimize the error between the projected 3D points and their image locations. 

Several works have underlined the importance of using uniformly distributed points in  sufficient amount, covering the different parts of the image and the whole altitude range of the scene \cite{tao2001comprehensive, long2015rpc}. As a result, regularized least squares methods have become widely used to gain robustness to different configurations and enforce well-conditioned normal equations \cite{tao2001comprehensive, long2015rpc, zhang2011rational, wang2016rpc}. Additionally, terrain-dependent strategies have explored the selection of optimal and balanced subsets of GCPs, e.g. \cite{tao2001comprehensive} propose a bucketing strategy or \cite{wang2016rpc} study the benefits of encouraging correspondences located at building edges in urban scenarios. In contrast, terrain-independent methods can arbitrarily generate regular sets of points, but require special care to the boundaries and density of the structure, e.g. \cite{zhang2011rational} investigate the impact of different number of elevation layers for flat and mountainous areas. 
\section{Method}
\label{sec:method}

We follow a terrain-independent approach similar to \cite{tao2001comprehensive} to fit an RPC model to another input geolocation model. The data used to fit the model consists of a 3D grid of uniformly distributed Control Points (CNPs) within some longitude, latitude and altitude boundaries (Fig.~\ref{fig:teaser}). The 2D image point of each CNP can be obtained by projecting it with the input geolocation model, so that each sample results in 5 normalized values, i.e. $(X_i, Y_i,Z_i,r_i,c_i)$. For simplicity, we drop the subscript $n$ of normalized coordinates and replace it by $i$ to refer to the sample index.

Using $N$ CNPs, Equation~\ref{eq:rpc_projection} can be rewritten as a system of equations, in matrix form, following the derivation of~\cite{tao2001comprehensive}:
\begin{equation}
 WTI - WG = 0, 
 \label{eq:eq_system}
\end{equation}
where
\begin{equation*}
    \begin{aligned}
    W =& \, \text{diag} \scalemath{0.85}{ \left[ \frac{1}{b(\bm{X}_1)} , ...,  \frac{1}{b(\bm{X}_N)} , \frac{1}{f(\bm{X}_1)} , ..., \frac{1}{f(\bm{X}_N)} \right] }  \\
    & b(\bm{X}_i) = b(X_i, Y_i, Z_i), \\
    T =& \, \text{block diag} \left[ M_r, M_c \right] \quad M_r, M_c \in \mathbf{R}^{N \times 39}
    \end{aligned}
\end{equation*} 

\begin{equation*}
    \begin{aligned}
    M_{r_i} =& \left[ 1, Z_i, Y_i, ..., X^3_i , -r_i Z_i, -r_i Y_i, ..., -r_i X^3_i  \right]\\
    M_{c_i} =& \left[ 1, Z_i, Y_i, ..., X^3_i , -c_i Z_i, -c_i Y_i, ..., -c_i X^3_i  \right], \\
    I =& \left[ a_0 , ... , a_{19} , b_1 , ... , b_{19}, e_0 , ... , e_{19} , f_1 , ... , f_{19}\right]^T, \\
    G =& \left[ r_0 , ... , r_N , c_0 , ... , c_N \right]^T.
     \end{aligned}
\end{equation*}   
In Equation~\ref{eq:eq_system}, $W$ is a weight matrix with shape $2N \times 2N$, $b(\bm{X}_i)$ denotes the RPC polynomial $b$ evaluated with the 3D coordinates of the $i$-th CNP; $T$ is the design matrix with shape $2N \times 78$; $I$ is the solution vector with the 78 RPC coefficients necessary to determine $a$, $b$, $e$ and $f$ in \mbox{Equation~\ref{eq:rpc_projection}}; and $G$ is a vector of length $2N$ containing the CNPs image coordinates.

Equation~\ref{eq:eq_system} can be solved by least squares minimization to estimate $I$, using the normal equation
\begin{equation}
 T^T W^2TI - T^T W^2G = 0.  
 \label{eq:normal_equation}
\end{equation}
To increase numerical stability, ridge estimation regularization \cite{wang2016rpc, zhang2011rational} is often added so that the normal equation becomes 
\begin{equation}
     (T^T W^2T + h^2 E ) I - T^T W^2G = 0,
    \label{eq:regu_normal_equation}
\end{equation} 
where $E$ is the identity matrix and $h$ is a scalar controlling the regularization that is applied. To choose the best regularization factor $h$, the L-curve criterion was introduced in \cite{zhang2011rational}. This heuristic computes the log norm of the solution $\left( \operatorname{log} \left\| I \right\|_{h} \right)$ versus the log norm of the residual $\left( \operatorname{log} \left\| WTI - WG \right\|_{h} \right)$ across different values of $h$ that extend from the minimal to the maximal singular value of T. This curve usually has a \mbox{L-shape}, in which the optimum corresponds to the maximum regularization parameter that achieves a small residual. The value corresponding to the corner of the curve, at the position of maximal curvature, is taken to set $h$   automatically.
For non-weighted regularized least squares (i.e. weights are set to the identity, $W = E$), the L-curve criterion is fast since the curvature can be computed with closed form expressions \cite{hansen1993use}. 

Therefore, we first set $W^{(0)} = E$ (where the superscript denotes the iteration number) and use the L-curve criterion to determine the optimal $h$ and an initial solution $I^{(0)}$. Then, for $i \ge 1$ , $W^{(i)}$ is determined from $I^{(i-1)}$ and is plugged in Equation \ref{eq:regu_normal_equation} to solve for $I^{(i)}$ iteratively (the SVD least squares solver is used for stability). The iterations stop when the change in terms of RMSE between the RPC projected CNPs and their image coordinates  becomes lower than a tolerance value. After convergence some final ICCV (Iteration by Correcting Characteristic Value \cite{zhang2011rational}) iterations are computed to remove possible biases introduced by the regularization. The same stopping criterion based on the RMSE improvement is used. Each ICCV iteration $k$ can be expressed as
\begin{equation}
(T^T (W^{(k)})^2 T  + E)I^{(k)}  =  T^T(W^{(k)})^2 G + I^{(k-1)}.
\label{eq:ICCV}
\end{equation}
\section{Experiments}
\label{sec:experiments}

\subsection{Data and use cases description}
\label{sec:data}

\begin{itemize}[leftmargin=*]
    \item {\bf{SAR.}} 32 Sentinel-1 SAR images (Table~\ref{tab:datasets}). The data is in Interferometric Wide Swath (IW) mode, each product contains 3 subswaths, and each subswath contains multiple bursts that need to be stitched together to get a continuous image. We construct the Range-Doppler physical sensor model and use the method from Section~\ref{sec:method} to fit an equivalent RPC model for each image of the dataset.
    
    \item {\bf{Optical.}} 47 \mbox{WorldView-3} panchromatic images (Table~\ref{tab:datasets}), from the 2016 \textit{IARPA Multi-View Stereo 3D Mapping Challenge} \cite{bosch2016multiple}. The original RPCs of the images exhibit small inaccuracies, mainly due to inexact knowledge of the satellite attitude angles, which cause a 3D point to be projected to non corresponding pixels across different images. Bundle adjustment (BA) algorithms are a well-known approach to correct RPC errors \cite{fraser2006sensor, beyer2018ames, mari2019bundle}. 
We apply a BA similar to \cite{beyer2018ames} to correc the projection function $\mathcal{P}$ of each RPC into a new $\mathcal{P}_{\text{BA}}$, expressed as
\begin{equation}
      \mathcal{P}_{\text{BA}}(X) = \mathcal{P} (R (X - T - C) + C),
     \label{eq:optical_rpc_correction}
\end{equation}
where $X$ is a 3D point. That is, each RPC is corrected by applying a translation $T$ followed by a rotation $R$ around an approximate camera center $C$, before applying the original projection $\mathcal{P}$. $C$ is derived by regressing a projective model from each  RPC model. 
Our method from Section~\ref{sec:method} is used to fit $\mathcal{P}_{\text{BA}}$ from the composition of  $\mathcal{P}$ with $T, R$, $C$.
\end{itemize}

\begin{table}[t]
\centering
\footnotesize
\begin{tabular}{ c | c c  } 
  & {\textbf{SAR dataset}} & {\textbf{Optical dataset}} \\   \hline
  platform &  S1 A/B  & WorldView-3\\ \hline
  number of images & 32 & 47 \\ \hline
  geographic area & Albania & Argentina\\
  (lon, lat) center &  (18.82,41.02) & (-58.61,-34.47) \\  \hline
  first acquisition date & 2019-08-03 & 2014-11-15 \\ \hline
  last acquisition date & 2020-02-05 & 2016-01-13 \\ \hline
   altitude range (m) & $\left[-533, 2969\right]$ & $\left[-513, 548\right]$ \\
 \end{tabular}
\caption{SAR and optical data used in the experiments. The altitude range of the area covered by each collection of images is defined using the [min, max] values from the corresponding SRTM digital elevation model $\pm 500$ m to consider tall buildings or fine irregularities beyond bare ground level.}
\label{tab:datasets}
\end{table}

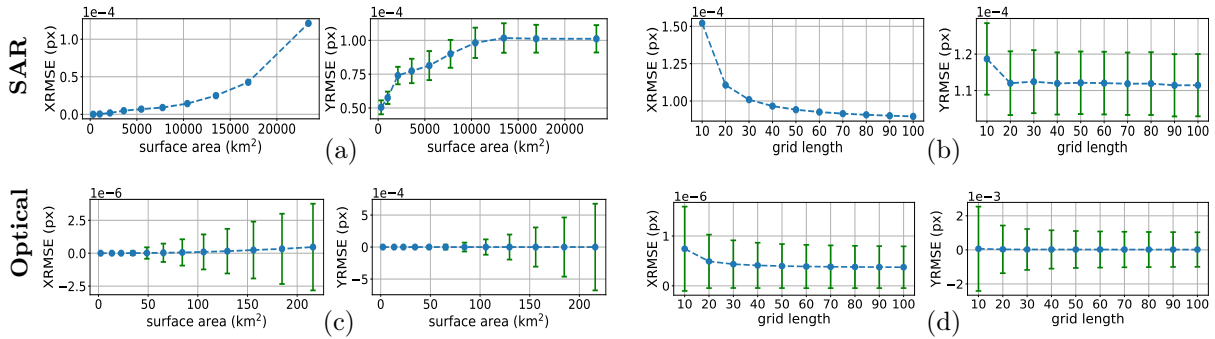
\begin{figure*}[t]
   \begin{tabular}{c c}
     \begin{tikzpicture}
     \node[anchor=south west, inner sep=0, outer sep=0] (img) at (0,0){
    \includegraphics[width=0.46\textwidth, height=2.1cm]{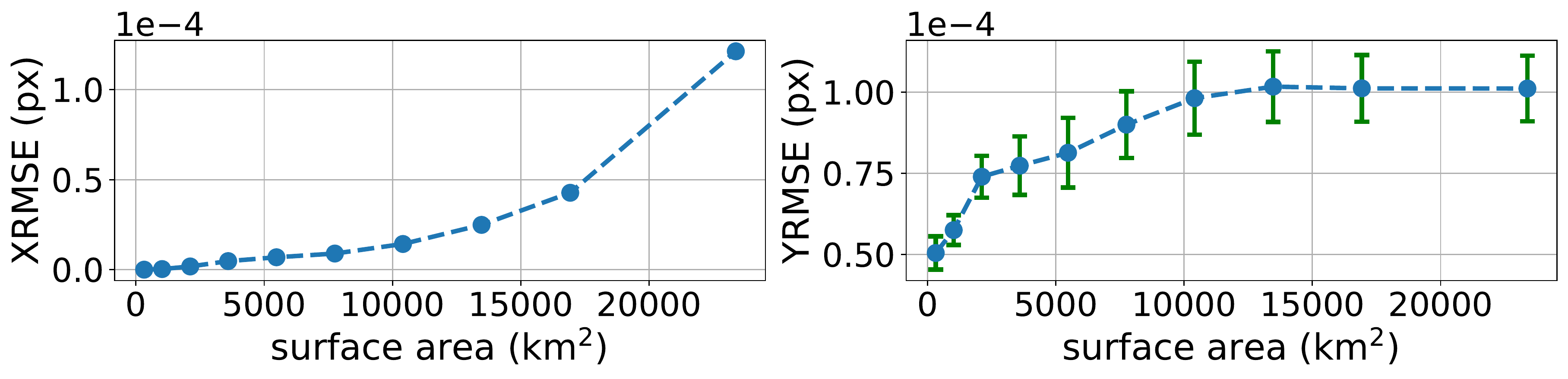}};
    \begin{scope}[x={(img.south east)},y={(img.north west)}]
       \node[text width=0.5cm] at (0.53,0.05) {(a)};
       \node[text width=1cm, rotate=90] at (-0.04,0.65) {\small \bf{SAR}};
    \end{scope}
    \end{tikzpicture} &
     \begin{tikzpicture}
     \node[anchor=south west, inner sep=0, outer sep=0] (img) at (0,0){
    \includegraphics[width =0.46\textwidth, height=2.1cm]{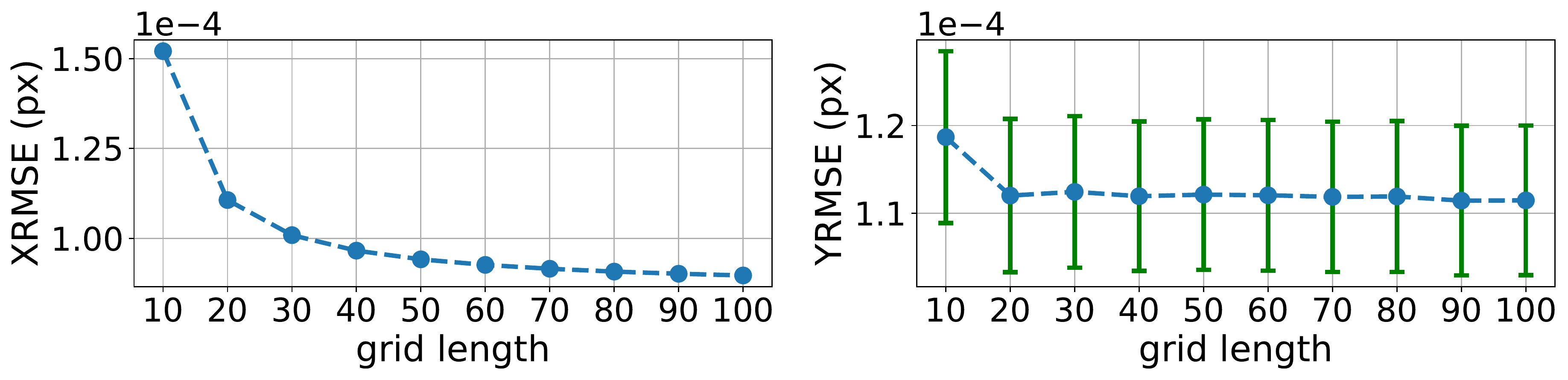}};
    \begin{scope}[x={(img.south east)},y={(img.north west)}]
       \node[text width=0.5cm] at (0.53,0.05) {(b)};
    \end{scope}
      \end{tikzpicture} \\ 
        \begin{tikzpicture}
     \node[anchor=south west, inner sep=0] (img) at (0,-0.1cm){
    \includegraphics[width=0.46\textwidth, height=2cm]{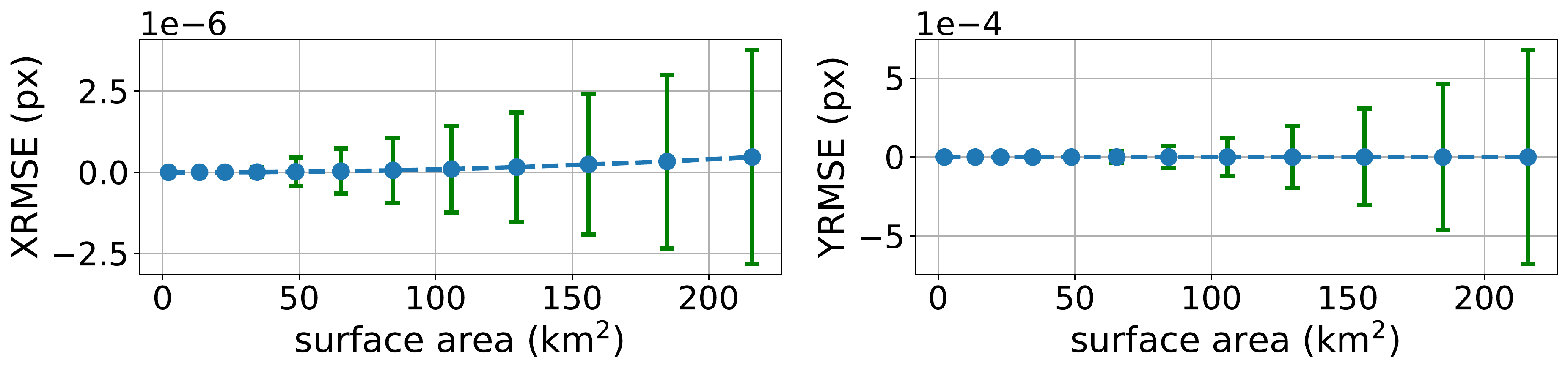}};
    \begin{scope}[x={(img.south east)},y={(img.north west)}]
       \node[text width=0.5cm] at (0.53,0.05) {(c)};
       \node[text width=1cm, rotate=90] at (-0.03,0.6) {\small \bf{Optical}};
    \end{scope}
    \end{tikzpicture} &
     \begin{tikzpicture}
     \node[anchor=south west, inner sep=0] (img) at (0,0){
    \includegraphics[width=0.46\textwidth]{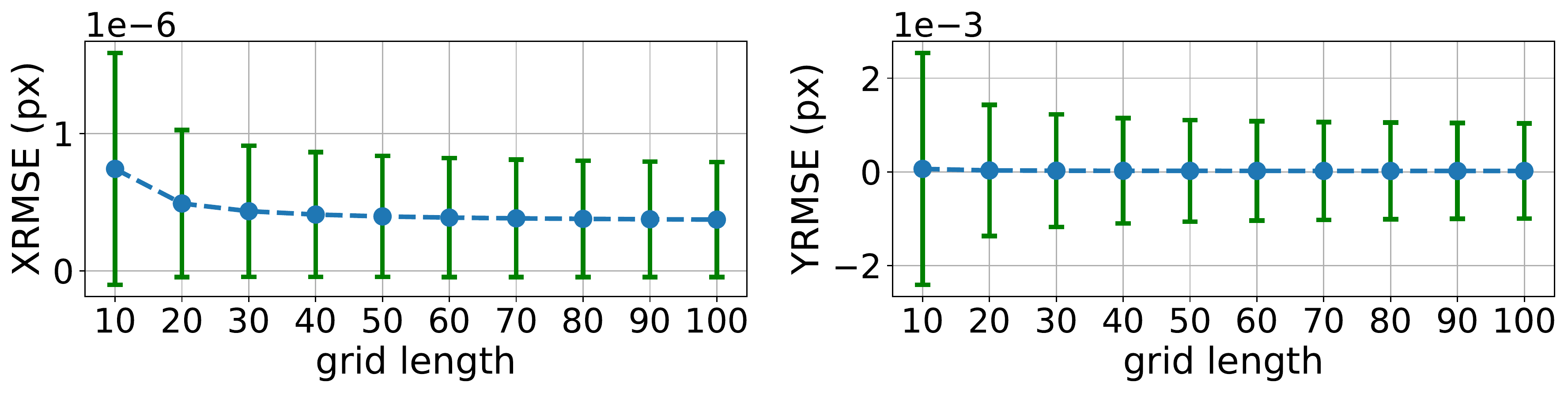}};
    \begin{scope}[x={(img.south east)},y={(img.north west)}]
       \node[text width=0.5cm] at (0.53,0.05) {(d)};
    \end{scope}
      \end{tikzpicture}   
      \end{tabular}
    \caption{RPC fitting error varying grid length and surface area on the SAR (a-b) and optical data (c-d) described in Section~\ref{sec:data}. Each vertical bar corresponds to the $[-\sigma/2, \, \mu, \, \sigma/2]$ values of the root-mean-square error (RMSE) evaluated across the different images of the datasets, in each dimension of the image plane, where $\mu$ corresponds to the mean and $\sigma$ to the standard deviation.}
    \label{fig:results}
\end{figure*}

\subsection{Performance assessment}

To apply the method described in Section~\ref{sec:method}, the altitude limits of the grid of CNPs are set using the altitude ranges in Table~\ref{tab:datasets}. To assess the RPC fitting, we use a grid of Check Points (CKPs), which are located in the middle of each pair of consecutive CNPs. The RMSE between the image coordinates obtained by projecting the CKPs with the output RPC and the image coordinates obtained using the input geolocation model, measured in pixels, is used as evaluation metric. We set a tolerance of $10^{-10}$ for the stopping criterion based on the RMSE improvement, and a maximum of $20$ iterations for the weighted least squares and the ICCV iterations. 

Two type of experiments were conducted to assess the performance and robustness of the method:
\begin{itemize}
    \item [-] \textit{Varying surface area}. For each image, we fit different RPCs by gradually increasing the longitude and latitude limits of the grid of CNPs from a small square centered at the image center to a larger square including the entire image. The number of CNPs is fixed, with $50 \times 50$ samples in the longitude and latitude dimensions and 10 elevation layers (25000 points in total).
    \item [-] \textit{Varying grid length}. For each image, we fit different RPCs, by increasing the number of CNPs in each elevation layer, i.e. $n \times n$ where $n$ is the \textit{grid length}. The longitude and latitude boundaries of the grid are fixed using the equivalent limits of the image plane. 
\end{itemize}
Overall, the results presented in Fig.~\ref{fig:results} show that the RPCs constructed with our method approximate the geolocation models with very high accuracy for the two datasets and the two scenarios outlined in Section~\ref{sec:data}. The RMSE is in the order of $10^{-4}$ pixels or less in both dimensions of the image plane for the majority of configurations that were tested, which emphasizes the robustness of the method. 

The experiments with different grid lengths (Fig.~\ref{fig:results}b and \ref{fig:results}d) show that 10 samples in the longitude and latitude directions is already a good choice and increasing this number beyond 20 does not result in significant improvements. The experiments with varying surface area (Fig.~\ref{fig:results}a and \ref{fig:results}c) show that both the overall RMSE values and its variation across the different images increase  with the size of the area being fitted. This is probably due to the fact that the SAR physical sensor model is less smooth for large neighborhoods. A similar behavior is obtained with the optical dataset, where the original RPCs are known to behave locally as an affine camera \cite{fraser2006sensor, mari2019bundle}. 
\section{Conclusion}
\label{sec:conclusion}

This article described an automatic algorithm to fit the RPC model of a satellite image in a terrain-independent manner. The inputs of the  method are a regular grid of 3D points (CNPs), with multiple elevation layers, and the 2D locations of the points on the image plane. We evaluated the method on real scenarios using collections of SAR and optical satellite images, and assessed its performance by varying the CNPs configuration. Finally, we release an open-source implementation of the algorithm as an easy-to-use Python package.

\section*{Acknowledgements}
Work partly financed by IDEX Paris-Saclay IDI 2016, ANR-11-IDEX-0003-02, Office of Naval research grant N00014-17-1-2552 and  N00014-20-S-B001, DGA Astrid project  \mbox{\guillemotleft \ filmer la Terre \guillemotright} n\textsuperscript{o}~ANR-17-ASTR-0013-01, MENRT,  and by a grant from Région Île-de-France.

\bibliographystyle{IEEEbib}
\bibliography{refs}

\end{document}